\documentclass[a4paper]{PoS}
\pdfoutput=1
\usepackage{bm,amsmath,amssymb,bbold}

\graphicspath{{./1_Figures/}}

\DeclareMathOperator\tr{tr}

\renewcommand{\Re}{\mathrm{Re}}
\newcommand{\der}{\partial}

\newcommand{\dd}{\mathrm{d}}
\newcommand{\bep}{\begin{pmatrix}} 
\newcommand{\eep}{\end{pmatrix}}

\newcommand{\SU}{\text{SU}}

\newcommand{\1}{\mathbb{1}}

\renewcommand{\epsilon}{\varepsilon}

\newcommand{\lat}{\text{lat}}

\def\ba#1\ea{\begin{align}#1\end{align}}
\def\akakko#1{\left\langle #1 \right\rangle}

\def\kkakko#1{\left[ #1 \right]}
\title{Lifshitz-type SU(N) lattice gauge theory in five dimensions}

\ShortTitle{Lifshitz-type $\SU(N)$ lattice gauge theory in five dimensions}

\author{\speaker{Takuya Kanazawa}\\
        iTHES Research Group and 
        Quantum Hadron Physics Laboratory,
        RIKEN, Wako, Saitama 351-0198, Japan
        \\
        E-mail: \email{takuya.kanazawa@riken.jp}}

\author{Arata Yamamoto\\
        Department of Physics, 
        The University of Tokyo, Tokyo 113-0033, Japan
        \\
        E-mail: \email{arayamamoto@nt.phys.s.u-tokyo.ac.jp}}

\abstract{We present a lattice formulation of 
non-Abelian Lifshitz-type gauge theories. Due to anisotropic 
scaling of space and time, the theory is asymptotically free even in five dimensions. 
We show results of Monte Carlo simulations that suggest  
a smooth approach to the continuum limit. }

\FullConference{The 33rd International Symposium on Lattice Field Theory
\\14 -18 July 2015\\Kobe International Conference Center, Kobe, Japan*}

\begin{document}

\section{Introduction}

Gauge theories exhibit a variety of phases 
depending on the dimensionality of spacetime. The critical 
dimensionality of Lorentz-invariant gauge theories is $3+1$, at which the 
coupling is classically marginal and quantum effects drive non-Abelian 
gauge theories with not-too-many fermion flavors 
to a strong-coupling confining phase with 
chiral symmetry breaking, whereas Abelian gauge theories hit 
a Landau pole at high energy and lacks a sensible definition 
at least within a perturbation theory.  
Gauge theories in lower dimensions are asymptotically free 
for any number of flavors and are relevant to strongly correlated 
lower-dimensional electronic materials. By contrast, at least within naive power 
counting, gauge theories living in higher than 4-dimensions lose 
renormalizability and seem to be ill-defined in the ultraviolet. However, some 
theoretical studies on physics beyond the Standard Model prompt considering 
gauge theories in higher dimensions, e.g. to solve the hierarchy problem 
in particle physics \cite{Randall:1999ee}.   
Given the lack of a continuum limit in the perturbative regime, one has to 
either conclude that a Yang-Mills theory in higher dimensions serves 
only as an infrared effective description of a more fundamental theory 
such as string theory, or, try to find their UV completion within quantum
field theory. The latter calls for an intrinsically nonperturbative approach. 

A powerful way to address nonperturbative problems in gauge theories 
is a lattice gauge theory pioneered by Wilson \cite{Wilson:1974sk}, in which one puts a theory 
on a discrete lattice keeping exact gauge symmetry. This proved to be 
a quite powerful numerical tool to elucidate the strong-coupling physics 
of Yang-Mills theory and QCD. Of course our ultimate interest is in 
physics in the continuum, so any lattice simulations of QCD have to be 
accompanied by a procedure of extrapolation to the continuum limit, 
which is quite well understood in QCD; it is the large-$\beta$ limit that 
corresponds to the domain of infinitely long correlation length in lattice units. 
By contrast, the existence of a continuum limit is far from trivial in 
unconventional theories. In the case of five-dimensional Yang-Mills 
theories on a lattice, the phase diagram consists of a weakly coupled 
Coulomb phase and a strongly coupled confining phase, which are 
separated by a first-order phase transition \cite{Creutz:1979dw}; 
so far the existence of a continuum limit is not confirmed yet.  
Although the so-called \emph{layer phase} proposed by Fu and Nielsen \cite{Fu:1983ei} 
has been actively investigated with lattice simulations, the issue of continuum limit 
is still under a debate \cite{Farakos:2010ie,Knechtli:2011gq,DelDebbio:2013rka}. 

There is however a class of theories that enjoy Lifshitz-type scale 
invariance instead of Lorentz invariance. The original motivation for them 
came from anisotropic 
critical points in condensed matter systems, where the critical scaling 
exponents are spatially anisotropic \cite{Chaikin_book}. 
In elementary particle physics, 
the idea of exploiting anisotropic scaling of space and time as a way of ameliorating 
UV divergences was explored by Ho\v{r}ava for quantum gravity 
\cite{Horava:2009uw} and by many authors for scalar and gauge theories, 
as reviewed in \cite{Alexandre:2011kr}. 
Although this deformed class of gauge theories seems to 
provide an intriguing UV completion, 
most of the analyses so far has been done in the continuum 
within a perturbative framework. 

In this work, we report on a first nonperturbative 
lattice regularization of higher-dimensional 
non-Abelian Lifshitz-type gauge theories \cite{Kanazawa:2014fla}. 
Besides a theoretical proposal 
of the formulation, we also performed a lattice 
Monte Carlo simulation based on our lattice action with the 
gauge group $\SU(3)$ and found a smooth 
crossover from strong to weak coupling. This opens up a novel arena 
to test ideas related to the beyond-Standard-Model physics 
and to seek for new dynamics of gauge fields unseen 
in the conventional setup in 4 dimensions.

\section{Lattice formulation}

We consider a Lifshitz-type gauge theory in $(D+1)$-dimensional 
Euclidean spacetime studied by Ho\v{r}ava \cite{Horava:2008jf}. The action is given by  
\ba
	S = \frac{1}{2} \int \dd x_0 \dd^D x 
	\kkakko{
		\frac{1}{e^2}\tr F_{0i}^2 + \frac{1}{g^2}\tr(D_i F_{ik})(D_j F_{jk})
	}\,, 
	\label{eq:S}
\ea
where $F_{\mu\nu} \equiv \der_\mu A_\nu - \der_\nu A_\mu + i [A_\mu, A_\nu]$ 
and $D_i F_{jk} \equiv \der_i F_{jk} + i [A_i, F_{jk}]$ ($i,j,k\in\{1,2,\dots,D\}$) 
with $A_\mu=A_\mu^a T^a$ the gauge field in the Lie algebra of $\SU(N)$.  
While the temporal part of the action is second order in derivatives, the spatial part is fourth order, 
hence this action has the dynamical critical exponent $z=2$.  
The absence of the usual spatial kinetic term $\tr F_{ij}^2$ at first sight appears to be 
vulnerable to quantum corrections, but it has been shown in the stochastic quantization 
of Yang-Mills theory \cite{ZinnJustin:1986eq,Okano:1986vr} 
that $\tr F_{ij}^2$ is not engendered under renormalization. 
By the same token, other terms of the same scaling dimension (e.g., $\tr(F^3)$) 
are not generated either. This simplifying property makes 
\eqref{eq:S} an attractive testbed for general Lifshitz-type gauge theories. 

The two couplings $e$ and $g$ are independent at the classical level. 
Their $\beta$ functions at one loop 
indicate that this theory is asymptotically free for $D\leq 4$ \cite{Horava:2008jf}. 
For $D=4$, a dynamical scale 
is expected to emerge in IR due to dimensional transmutation, much like in QCD. 
This attractive conjecture due to Ho\v{r}ava cannot be tested in a perturbative framework, 
however. 

\begin{figure}[t]
	\centering 
	\includegraphics[width=.4\textwidth]{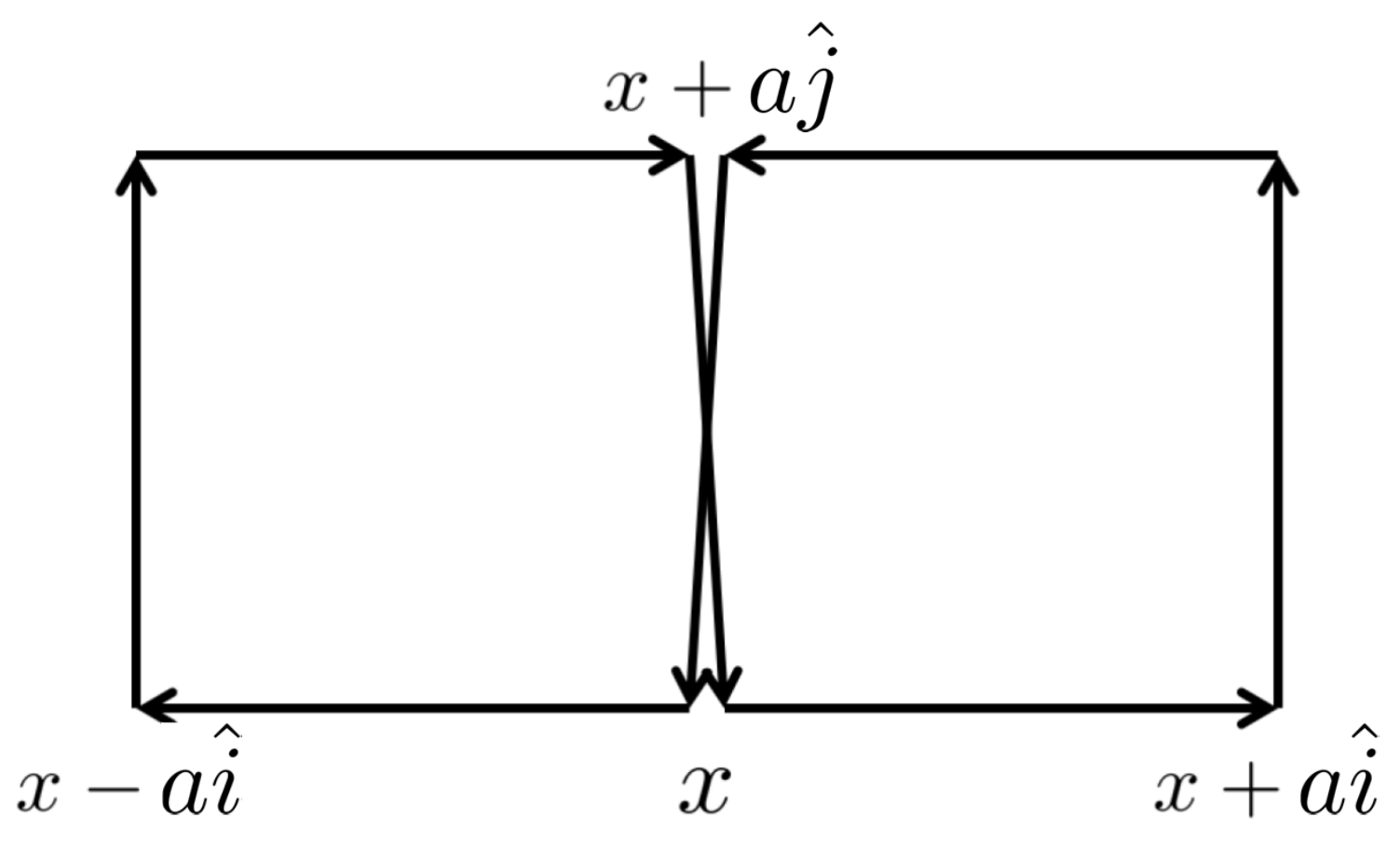}
	\caption{Twisted Wilson loop $T_{ij}(x)$ that starts and ends at a point $x$ 
	on the lattice. The variables $\hat{i}$ and $\hat{j}$ are unit vectors in $x^{i}$ and 
	$x^j$ directions, respectively. Note that $T_{ij}\ne T_{ji}$. 
	Figure taken from \cite{Kanazawa:2014fla}.}
	\label{fg:T}
\end{figure}

Now we proceed to a lattice regularization of \eqref{eq:S}. 
It is crucial to preserve gauge invariance at finite lattice spacing. We propose 
\ba
	S_{\lat} & = \frac{\beta_e}{2N}\sum_x \sum_{i=1}^D 
	\Re\tr\kkakko{\1-P_{0i}(x)} + 
	\frac{\beta_g}{2N}\sum_x \sum_{j=1}^D 
	\Re\tr\Big[ 
		\1-\prod_{\substack{i=1\\i\ne j}}^D T_{ij}(x)
	\Big]\,. 
	\label{eq:Slat}
\ea
The temporal part of $S_\lat$ uses the standard Wilson's $1\times 1$ 
plaquette $P_{0i}$ lying in the $(x^0,x^i)$-plane, whereas $T_{ij}$ 
in the spatial part is a twisted $2\times 1$ Wilson loop in the $(x^i,x^j)$-plane, 
as shown in Fig.~\ref{fg:T}. The order of multiplication in $\prod T_{ij}$ 
is arbitrary because it only affects irrelevant terms in the continuum limit.   
%%%%%%%%%%%%%%%%%%%%%%%
%%%%%%%%%%%%%%%%%%%%%%%
%%%%%%%%%%%%%%%%%%%%%%%
%%%%%%%%%%%%%%%%%%%%%%%
The lattice couplings $(e_\lat,g_\lat)$ are defined through 
\ba
	\beta_e=\frac{2N}{e_\lat^2} \quad \text{and} \quad  
	\beta_g=\frac{2N}{g_\lat^2}\,.
\ea
It can be straightforwardly checked that $S_{\lat}$ recovers $S$ 
in \eqref{eq:S} at least in the \emph{classical} continuum limit. On the lattice, 
we need to control the dimensionless couplings $\beta_{e,g}$ to find 
a second-order phase transition 
where the correlation length measured in lattice units tends to infinity. 
As shown in \cite{Horava:2008jf}, the physical couplings $(e,g)$ flow to 
zero in UV (asymptotic freedom), 
which motivates us to expect that the continuum limit of the lattice theory 
\eqref{eq:Slat} may be defined by the weak-coupling limit $e_\lat,g_\lat \to0$. 
We check this anticipation in full numerical simulations in the next section.

\section{Monte Carlo simulation}

\begin{figure}[t]
	\centering 
	\includegraphics[width=.55\textwidth]{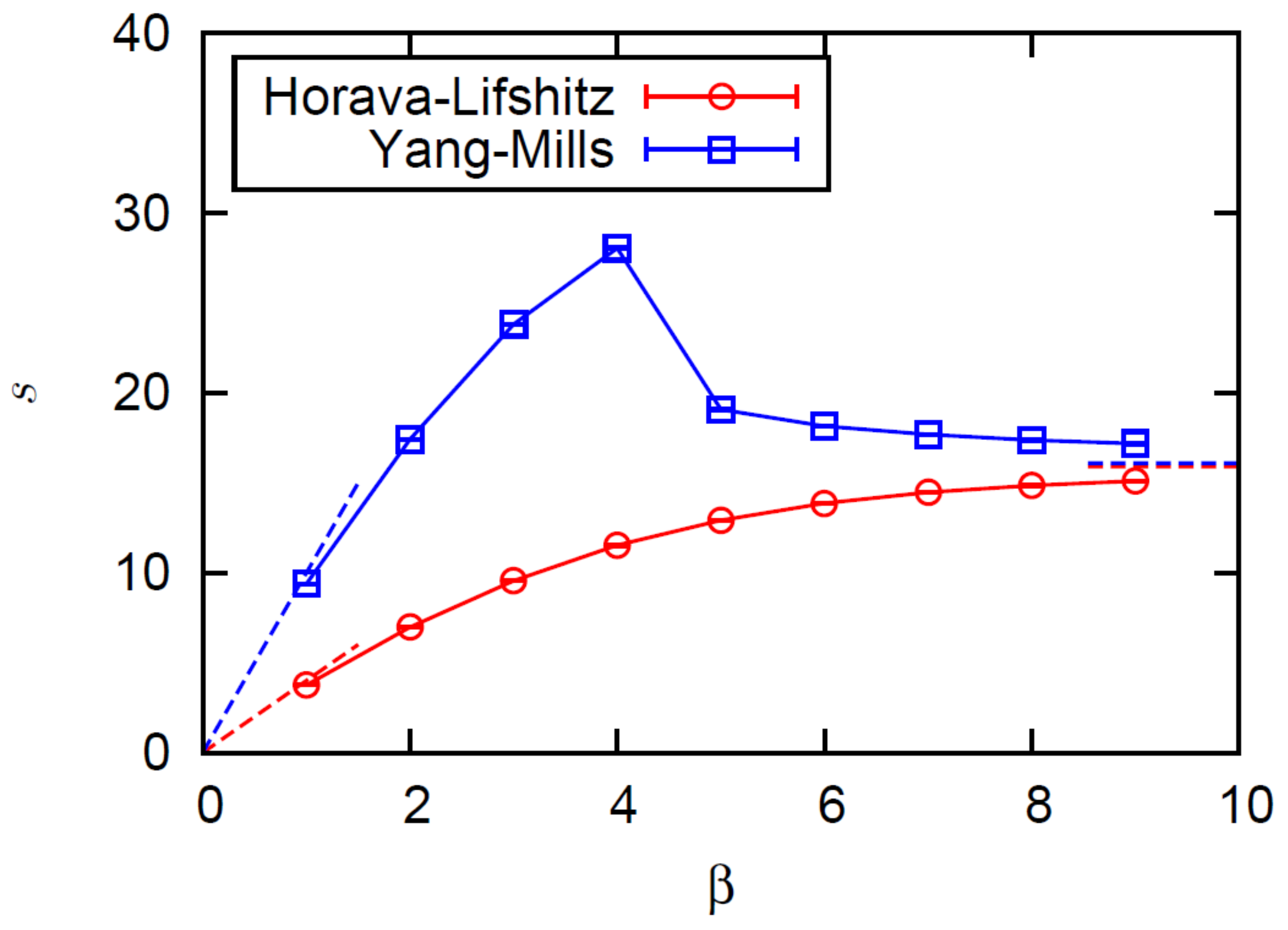}
	\caption{Action density $s\equiv \akakko{S_\lat}/N_\lat$ 
	with isotropic couplings $\beta_e=\beta_g\equiv\beta$ on a $6^5$ lattice 
	(red points). For comparison, simulation results for isotropic Yang-Mills theory on the 
	same lattice is plotted (blue points). Dashed lines represent the leading order of 
	weak and strong coupling expansions, respectively. 
	Figure taken from \cite{Kanazawa:2014fla}.}
	\label{fg:s}
\end{figure}

We simulated the lattice theory \eqref{eq:Slat} with standard Monte Carlo techniques. 
We studied the $N=3$ theory on lattices of size $N_\lat=6^5$ and $10^5$.  
In Fig.~\ref{fg:s} the action density for isotropic couplings is plotted. 
%%%%%%%%%%%%%%%%%%%%%%%
%%%%%%%%%%%%%%%%%%%%%%%
%%%%%%%%%%%%%%%%%%%%%%%
%%%%%%%%%%%%%%%%%%%%%%%
The transition from strong to weak coupling appears 
to be a smooth crossover. At $\beta\ll1$ and $\gg1$ 
one can estimate the behavior of the action density 
analytically and our numerical results are in full accord with 
those limits. For the purpose of comparison we have also 
simulated the conventional Yang-Mills theory with Wilson 
plaquette action on the same lattice, as plotted in blue in 
Fig.~\ref{fg:s}. There is a sharp drop at $\beta\sim 4.5$, 
which is a known bulk transition from a confining phase 
at small $\beta$ to a deconfined Coulomb phase at large $\beta$ 
\cite{Itou:2014iya}. This is consistent with perturbative non-renormalizability of 
the Yang-Mills theory in five dimensions. (A similar result was 
obtained for the gauge group $\SU(2)$ long time ago \cite{Creutz:1979dw}; 
for a more recent study, see \cite{deForcrand:2010be}.) 
Confirmation of this dramatic disparity between the two theories is 
the main result of this work. 

%%%%%%%%%%%%%%%%%%%%%%%
%%%%%%%%%%%%%%%%%%%%%%%
\begin{figure}[h]
	\centering 
	\includegraphics[width=.55\textwidth]{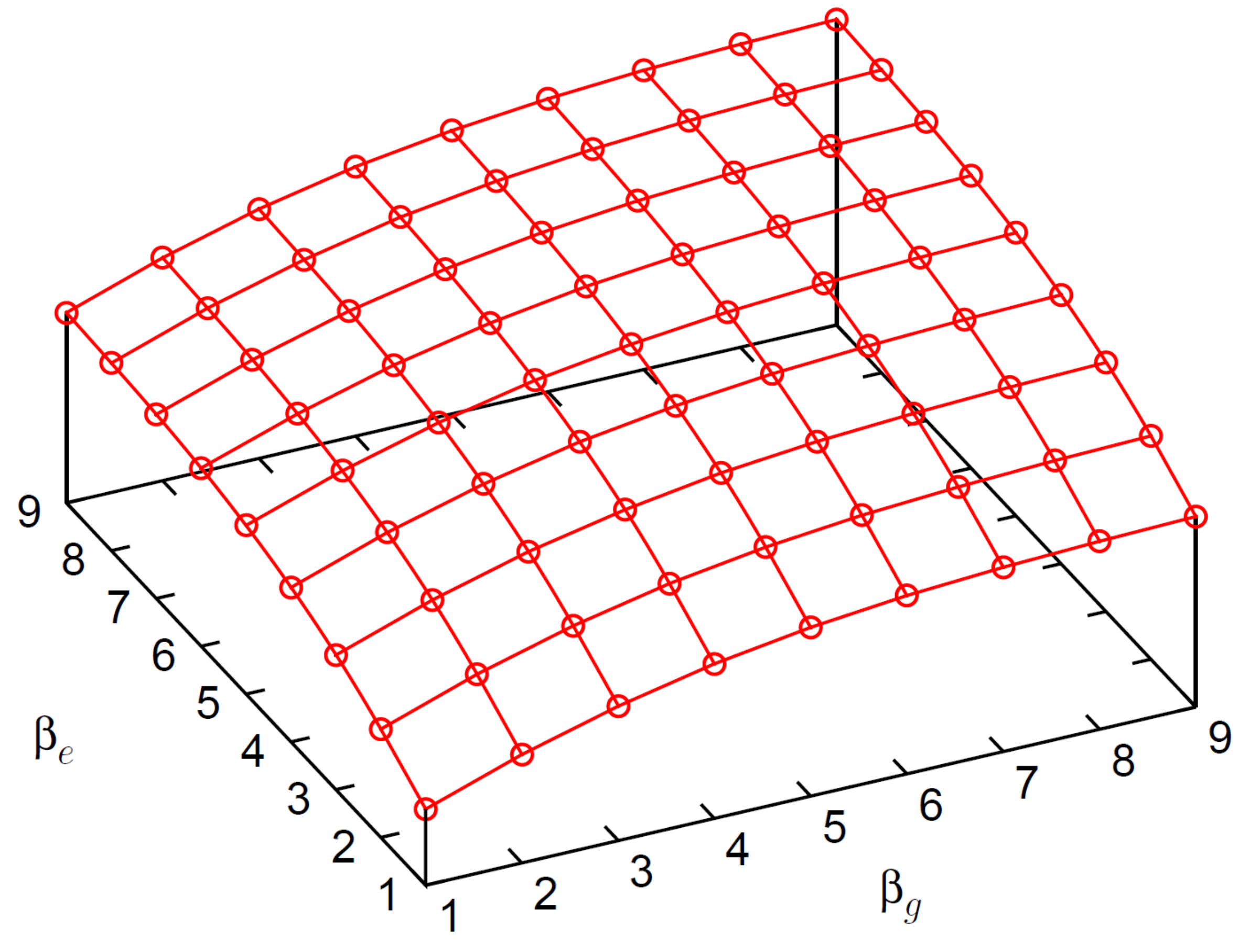}
	\caption{Action density on a $6^5$ lattice for various $\beta_e$ 
	and $\beta_g$. Figure taken from \cite{Kanazawa:2014fla}.}
	\label{fg:s2}
\end{figure}
%%%%%%%%%%%%%%%%%%%%%%%
%%%%%%%%%%%%%%%%%%%%%%%
We also performed simulations with anisotropic couplings  
($\beta_e\ne \beta_g$). The action density, plotted in Fig.~\ref{fg:s2},  
varies smoothly with the couplings and shows no indication of a phase transition.  

\begin{figure}[h]
	\centering 
	\includegraphics[width=.55\textwidth]{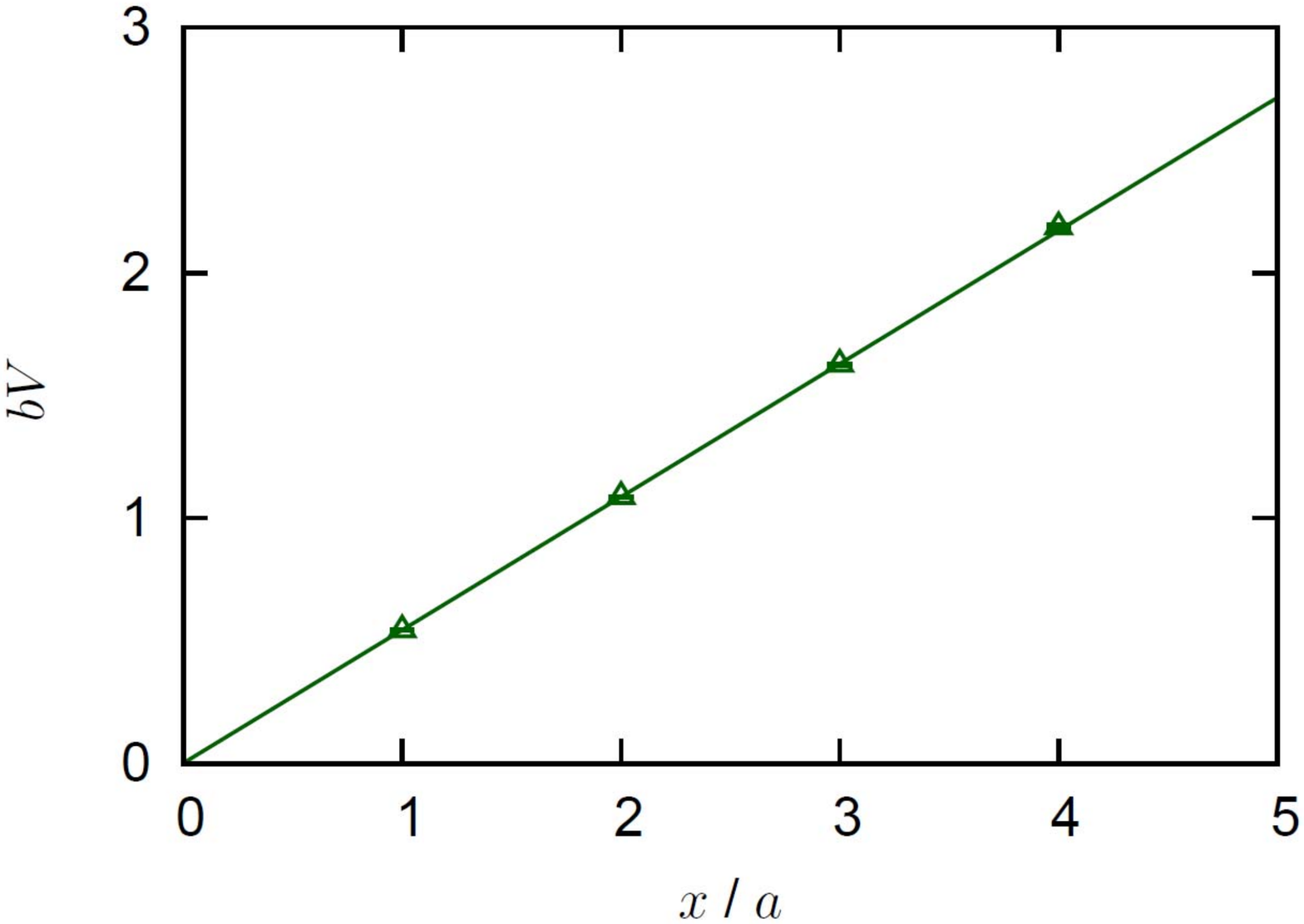}
	\caption{Color-singlet potential $V(x)$ measured on a $10^5$ lattice 
	with $\beta_e=\beta_g=9$. The axes are in lattice units. 
	Figure taken from \cite{Kanazawa:2014fla}.}
	\label{fg:V}
\end{figure}

Next we measured temporal Wilson loops $W_{0i}$ of edge lengths $t$ and $x$ 
to extract the color-singlet potential between infinitely heavy quarks 
\ba
	V(x) & \equiv - \lim_{t\to \infty} \frac{1}{t}\log\akakko{\tr W_{0i}(t,x)}. 
\ea
In actual simulations, the extrapolation to $t=\infty$ is replaced with 
a large but finite $t$. The numerical result for the potential is presented in Fig.~\ref{fg:V}. 
The data points indicate a linear confining potential between heavy quarks in this 
theory. Due to limited volume, we could not extract the potential at very 
short and long distances. In perturbation theory, one expects a one-gluon-exchange 
potential of the form $V(x)\sim \int \dd^D p\, \exp(-ipx)/p^2\sim 1/x^2$ at small $x$. 
Confirmation of this behavior is postponed to future work.  
%%%%%%%%%%%%%%%%%%%%%%%
%%%%%%%%%%%%%%%%%%%%%%%
%%%%%%%%%%%%%%%%%%%%%%%
%%%%%%%%%%%%%%%%%%%%%%%

We have also measured spatial Wilson loops $W_{ij}$ and spatial plaquettes 
$P_{ij}$.  Their statistical averages were found to be zero within errors. This is natural, considering 
that the spatial kinetic term for gluons $\tr F_{ij}^2$ should not be generated by quantum 
corrections.

\section{Conclusions}

In this work we reported a nonperturbative lattice regularization of Lifshitz-type 
gauge theories with anisotropic scale invariance. The results of first numerical simulation 
of our lattice theory in 5 dimensions reported here suggests  
that the continuum limit can be defined in the weak-coupling limit, in a way 
quite analogous to the standard lattice Yang-Mills theory in 4 dimensions. 
We measured the heavy-quark potential and found that quark confinement takes place 
in our lattice theory even at weak coupling.  All these findings are theoretically interesting 
as a new pathway to extend the realm of cutoff-free gauge theories into higher 
dimensions. While our first simulation was limited to a relatively small volume 
and a restricted coupling parameter region, there seems to be no fundamental 
difficulty in performing more thorough simulations based on existing techniques 
in lattice QCD.   

There are numerous questions left unanswered in this pilot study. Can we extend 
the lattice formulation to the case of dynamical exponents $z>2$? What is the 
mechanism of quark confinement in 5 dimensions? Can we 
introduce quarks into the theory without spoiling good renormalization properties?  
If possible, then could there be a spontaneous flavor symmetry breaking of quarks? 
Is there any applications to the theories of 
quantum criticality with anisotropic scale invariance 
in condensed matter systems \cite{Ardonne:2003wa,Freedman:2004ki}? 
What can be learned about the Standard Model from this theory? 
These issues should be investigated elsewhere.

\acknowledgments

TK was supported by the RIKEN iTHES Project. 
AY was supported by JSPS KAKENHI Grants Number 15K17624.
The numerical simulations were performed by using the RIKEN 
Integrated Cluster of Clusters (RICC) facility.

%\bibliographystyle{JHEP}
%\bibliography{Horava_proceedings}
\bibliography{proc_anisotropic_LGT.bbl}
\end{document}